# Oxidation of $In_2S_3$ films to synthetize $In_2S_{3(1-x)}O_{3x}$ thin films as a buffer layer in solar cells

S. Laghrib[1], M. Hamici[1], Y. Gagou[2], L. Vaillant Roca[3], and P. Saint-Grégoire[4*]

(1) University Ferhat Abbas, Setif 1, ALGERIA

(2) LPMC, University of Picardie Jules Verne, 80039 Amiens cedex 01, FRANCE

(3) Enermat Division, Institute for Science & Technology of Materials, IMRE, Havana University, 10400 Havana, CUBA

(4) MIPA Laboratory, University of Nîmes, 30021 Nîmes cedex, FRANCE

*Corresponding author: pstgregoire@gmail.com

**Abstract:**

$In_2S_{3(1-x)}O_{3x}$ is known from preceding studies to have a bandgap varying continuously as a function of x, the reason why this solid solution is potentially interesting in the field of photovoltaics. In this work, we present results on oxidation of $In_2S_3$ by heating in air atmosphere to obtain the desired material. The oxidation is accompanied by a mass loss due to the substitution of S by O atoms that is studied by means of thermogravimetric analysis. It appears that the temperature region in which the oxidation occurs is strongly dependent on the texture of deposited films. As-grown films deposited by chemical bath deposition are subjected to nano-oxidation occurring at lower temperature than oxidation of materials that are characterized by a better crystallinity and larger crystallite size. X ray diffraction and scanning electron microscopy (including EDX) were used to get information on the compounds and the texture of films. The main conclusion of the paper opens the perspective of practical applications for producing layers for solar cells.

**Introduction**

$In_2S_3$ is a very versatile material that has been studied in the last years mainly as CdS buffer layer substitute for fabricating Cd-free CIGS solar cells [1, 2, 3]. One of its advantages is that it can be obtained from solution, using a non-expensive growing technique as Chemical Bath Deposition (CBD) [4, 5], in its β form [6]. In the literature the oxidation of $In_2S_3$ films was described for films obtained by physical vapor deposition (PVD) [7] and for films obtained by spray pyrolysis [8]. The main interest of this approach is that the band gap, that varies as a function of composition, can be tailored to intermediate values between the band gap value for $In_2S_3$ and that for $In_2O_3$ [7]. Increasing the band gap by inducing the oxidation of $In_2S_3$ may provide a decrease in absorption losses, leading to improving short circuit currents in solar cells. At the same time, it might provide a better match with the used transparent conducting oxide. This is important since this material is involved in promising solar cells, those which are built with kesterites, that present interesting perspectives of low cost solar cells, produced with semiconductors composed of abundant and cheap elements (Cu, Zn, Sn, S).

For applications, it is important to master the quality of the films, and it is why we performed the synthesis of $In_2S_{3(1-x)}O_{3x}$ films by oxidation of $In_2S_3$ films, and paid a particular attention to the structure of films, to the homogeneity of the chemical composition, to the oxidation process, and to structure modifications all along the oxidation process or thermal treatment, for films obtained using CBD.

In the following, we present results of X-ray diffraction (XRD) that was performed to check the composition of the original films, of thermogravimetric analysis (TGA) measurements done in order to get information on the mass variations versus temperature that are associated with the modification of chemical composition, and of scanning electron microscopy (SEM) that was used to obtain data on the morphology of the films and the composition.

**Experimental procedure**

$In_2S_3$ films were obtained by chemical bath deposition, varying conditions given in the protocol described in [9], in particular varying deposition time and concentrations of thioacetamide and $InCl_3$. The $In_2S_3$ film growth was performed on different glass substrates (including FTO glass) in a reactive solution, in a 100 ml beaker, and also on a pure silica substrate for the reason we shall present below. The preparation that was finally chosen after several tests coincides with that described in ref.9 ; we begin with the dissolution of

0.1 M InCl$_3$ in 10 ml of water, to which 20 ml of 0.5 M acetic acid and finally 20 ml of 1 M thioacetamide are added. The volume was finally completed with distilled water, to reach a total volume 100 ml and the preparation was heated till 70°C and continuously stirred. Deposition times ranged between 1 min and 60 min, and optically homogeneous yellow films of thickness up to approximately 600 nm were obtained.

In the following, the samples will be denoted as « F » for « films », and « FP » for the « film powder » that was extracted from the thin film by scrapping off from the deposited layer. As-grown samples, directly obtained from the bath without any other treatment than cleaning with distilled water and dried, were denoted with the prefix "ag", for instance "ag-FP" for film powder obtained by scrapping off from an as-grown deposited film. Powders extracted from films were used when the experimental techniques require to work with powders, for instance in the case of X-ray diffraction (XRD) with high enough accuracy to deduce reliable informations on the composition and structure, and TGA.

XRD was performed using a X-ray powder diffractometer Philips X-PERT PRO II working with CuKα1 and CuKα2 radiation. For layers on glass substrates (F samples), the incident angle was kept constant at a small value ($2\theta$ = 3°) whereas for powders we worked with classical Bragg Brentano scans. Data were analyzed using the Fullprof software [10], using the Rietveld method to check the nature of layers and study their composition.

Thermogravimetric measurements were performed by means of the instrument 2950 TGA (TA Instrument) in nitrogen or in air atmosphere, doing scans at constant temperature rate (« ramp mode »), generally at rates between 10°C/min and 20°C/min, or in isothermal mode. Scanning electron microscopy (SEM) was performed by means of Hitachi S4500 and S4800 instruments, in image mode and EDX.

**Experimental results**

The first check after syntheses has been done to verify that the deposited material is indeed indium sulfide in its β phase. Figure 1-a presents the X ray diffraction pattern collected with an as-grown In$_2$S$_3$ film denoted « ag-F ». The strongly varying background is mainly due to the glass substrate but seems to contain an amorphous component inherent to the layer. Comparing this pattern with JCPDS data Nr 73-1366, all reflexions are indexed, that allows us to identify the compound as being indeed the β– In$_2$S$_3$.

We collected also the diffraction pattern with as–grown powders (sample « ag-P ») and powders obtained from as-grown films ("ag-FP") which could also be shown to crystallize in the β– In$_2$S$_3$ structure. Because some authors [11] have mentioned the possibility that several phases were coexisting in the deposited material in their layer, we performed a profile adjustment starting from the known data for β–In$_2$S$_3$ (JCPDS data Nr. 73-1366) [12].

We found a good agreement (Fig. 1-b) with the presence of a single phase with the following refined parameters: a = b = 7.5612(9) Å, c = 32.0538(8) Å in $I4_1/amd$ space group (Nr 141). Results of Rietveld refinements both on lines position and on profile, is very satisfactory as can be seen in Fig. 1-b, where the calculated profile fits very well the diffraction data, and the difference between the experimental points and calculated curve led to a straight line near zero. The final reliability factors, $R_p$ = 4.97%, $R_{wp}$ = 6.27%, $R_{exp}$ = 5.40% and $\chi^2$ = 1.35 using Fullprof software, are satisfactory.

From XRD results, we do not have any evidence of another phase than that corresponding to the β– $In_2S_3$ structure in the layer and refinement results indicate a homogeneous composition. Having the confirmation that the deposited material was of homogeneous composition and was actually the compound with the β– $In_2S_3$ structure, we performed the following experiments:

- TGA measurements for runs in nitrogen atmosphere till 450°C, because this gas is expected not to interact chemically with the deposited layer
- TGA measurements in air atmosphere in as-grown samples and in samples after thermal run in nitrogen.

CBD deposited $In_2S_3$ thin films are known to contain hydroxides [1] and solvents still present as far as a thermal treatment has not been performed. TGA results of scans at constant rate in nitrogen atmosphere show mass losses (not shown in this paper) that we attributed to the departure of these hydroxides and solvents during heating up to 500°C : this assumption is supported by the observation that the sample mass on further cooling down to room temperature from 500°C, remains constant at the value reached at high temperature. Moreover, further scans at constant rate do not show any other loss at the corresponding temperatures.

Surprisingly, TGA results are very different for the as-grown sample (ag-FP) and the sample preliminary heated in nitrogen till 450°C (sample denoted as FP450N2).

We see in Fig. 2 the TGA result obtained with the powder ag-FP from the as – grown film. A mass loss occurs above 150°C and extends in a rather wide temperature range till approximately 380°C, above which the mass remains constant. In distinction with this curve, we see in Fig. 3 that when the same initial sample has been preliminarily submitted to a thermal treatment in an inert atmosphere ($N_2$), there occurs also a mass loss on heating in air atmosphere but it happens at higher temperature, around 500°C and in a smaller temperature interval (the mass loss begins in the neighborhood of 430°C and considerably slower down around 550°C). It is noteworthy that the temperature at which we observe oxidation in as – grown films is significantly lower than temperatures mentioned in [8] for samples deposited by spray pyrolysis, a fact that seems to confirm the sensitivity of oxidation temperature on sample texture.

We interpreted these results as due to differences in the crystallite size : as-grown samples are expected to be composed of nanometric grains and of amorphous regions, and consequently they can be expected to be more reactive and thus to interact chemically with $O_2$ at a lower temperature than samples having an improved crystalline quality (suppressed amorphous regions, increased grain size, and also probably a more homogeneous texture) due to annealing. To check this assumption we thereafter performed XRD measurements and SEM observation that will be described below.

XRD confirms that the final product is $In_2O_3$ and that the mass loss is due to the substitution of S by O atoms. XRD performed on samples collected after TGA thermal runs of figures 2 and 3, confirms indeed that the final product is pure $In_2O_3$. Figure 4 presents for instance the result obtained with sample after the TGA measurement of Fig.2 (as-grown FP sample heated in air atmosphere till 450°C). We obtained a very good agreement between calculated diagram for $In_2O_3$ lines and observed (experimental) ones, both on the position of lines and on their profile (see the inset in Fig.4).

The starting calculation was based on the known JCPDS Nr 06-0416 data for $In_2O_3$ [13]. The refinement converges rapidly according to the Rietveld method. The obtained lattice cell parameters, in the cubic space group *I* a -3 (Nr 206) are : a = b = c = 10.1203(9) Å. The obtained reliability factor in the profile adjustment mode ($R_p$= 5.57%, $R_{wp}$ = 7.37%, $R_{exp}$ = 5.41%, $\chi^2$ = 1.86) using Fullprof software, are very satisfactory .

On the other hand electron microscopy and EDX analysis confirm that thermal treatments performed in $N_2$ atmosphere do not change the chemical composition but change the texture. Moreover mention that the deduced composition of deposited layers is stoechiometric, with a proportion between In and S in the 2-3 ratio, as reported already for the same synthesis method by Sandoval Paz et al [ 9]. This observation was performed for samples deposited on soda-lime glass, on conducting 15 Ω/square FTO glass TCO22-15 of Solaronix, and also for one sample on pure silica. Results were identical in all cases so that an eventual change of texture due to diffusion of sodium from the substrate (known to be possible, see ref. [14]) is excluded, and the phenomenon is attributed exclusively to a thermal effect.

Figures 5a and 5b present respectively the texture of the as-grown film deposited on glass, and of the same type of film after annealing in nitrogen atmosphere at 450°C (heating rate of 10°C/min and annealing at 450°C during 15 min). The annealing leads to a coarsening of the texture and XRD data (not shown here) are characterized by sharper diffraction peaks, thus showing a better crystallinity. At the opposite, the as-grown layer has a nanometric texture, in which small nanocrystals have grown from the chemical bath.

Having the confirmation that the final product was $In_2O_3$ after the whole thermal process corresponding to Fig. 2 or 3, we attribute the mass variation (as shown in the curves of figures 2 and 3) entirely to the substitution of S by O, and further paid attention to the role of annealing in air atmosphere, at a temperature situated in the region where the mass loss occurs. Considering for instance the ag-FP sample that presents a mass loss between approximately 150°C and 380°C (see Fig.2), we performed with this sample a ramp from room temperature till 265°C and subsequent isothermal measurements (Fig.6). We observe that the mass loss goes on as a function of time, according to an exponential decay characterized by a relaxation time $\tau \sim 3,3$ min. However, we remark that the mass loss saturation value is smaller than that corresponding to complete oxidation deduced from Fig.2. We conclude that the oxidation process is uncomplete, even after long annealings in air atmosphere : for instance at 265°C, the mass loss at saturation corresponds to a deduced composition $In_2S_{3(1-x)}O_{3x}$ with $x \sim 0.3$, based on a linear extrapolation from mass variations accompanying the total oxidation, a conclusion that is supported by measurements of composition by EDX of the corresponding compound. Moreover, during the isothermal measurements, the variation of mass is small (from 0.0942 mg to 0.0914 mg – namely 2.97%), which shows that the curves obtained in the ramp mode (Figs 2 and 3) are close to curves that would correspond to equilibrium states, even at scan rates as high as 20°C/min. Finally, mention that the band gap that may be deduced from UV-visible spectra (not shown in this paper), is consistent with the deduced composition variation and earlier studies of ref [7]. Details will be shown elsewhere.

Performing EDX mapping we got the result that the layers obtained by such processes have a homogeneous chemical composition, a result consistent with the observation that XRD patterns of oxidized powders extracted from films do not correspond to a superposition of patterns (as evidenced in Rietveld analysis), but correspond to single phases. A careful crystallographic study of samples obtained at different annealing temperatures, should allow to determine if the substitution of S by O occurs with different probabilities in different crystallographic sites. Moreover the problem of the phase diagram is open, since there should occur at least one line in temperature-composition phase diagram, separating regions with different space groups ( *I* 4$_1$/*a m d* (Nr 141) on the S-rich side, and *I* a -3 (Nr 206) on the O-rich side). This work is in progress.

## Conclusion

The solid solution $In_2S_{3(1-x)}O_{3x}$ is known from literature [7] to have a band gap varying continuously as a function of x, which in principle presents the advantage in solar cells, to tailor it and adapt it to improve the characteristics of the device to which it is integrated. This idea could be particularly relevant in solar cells (where $In_2S_3$ is already used as buffer layer) as the promising photovoltaic cells built with kesterites that are the object of an intense attention nowadays [15, 16, 17]. In this paper, we show that the films we have grown by chemical bath deposition are constituted exclusively of β-$In_2S_3$. Surprisingly we obtained the important result that the thermal treatment to be performed in order to have a final composition and the corresponding band gap, depends very strongly on the sample morphology and particularly on crystallite size. Process temperatures are lower for as-grown samples, that are shown to possess an interesting nanostructure. Due to the presence of nanometric grains, they present a stronger reactivity in air : the (nano)oxidation occurs in these samples between 150°C and 380°C. At 380°C a complete oxidation occurs and the final composition of the film is confirmed to be $In_2O_3$. Besides the fundamental problem raised by this phenomenon, this situation presents potentially an important practical interest since process temperatures are low enough to avoid a damaging of other layers in the cell, and are compatible with the use of lime glass as substrate.  Temperature stabilization at intermediate values leads to uncomplete oxidation of the layer, namely to a layer with an intermediate composition $In_2S_{3(1-x)}O_{3x}$ with x between 0 and 1 depending on annealing temperature, a situation that opens the problem of the microscopic mechanisms leading to it.

**Acknowledgements**

Conseil Régional Languedoc-Roussillon for providing the financial support of invited professors

Embassy of France in Havana, Cuba, for financement of mobility of cuban researchers

The authors thank Bernard Fraïsse (ICGM, University Montpellier 2) for XRD measurements and Nathalie Masquelez (IEM, University Montpellier 2) for TGA measurements.

FIGURES

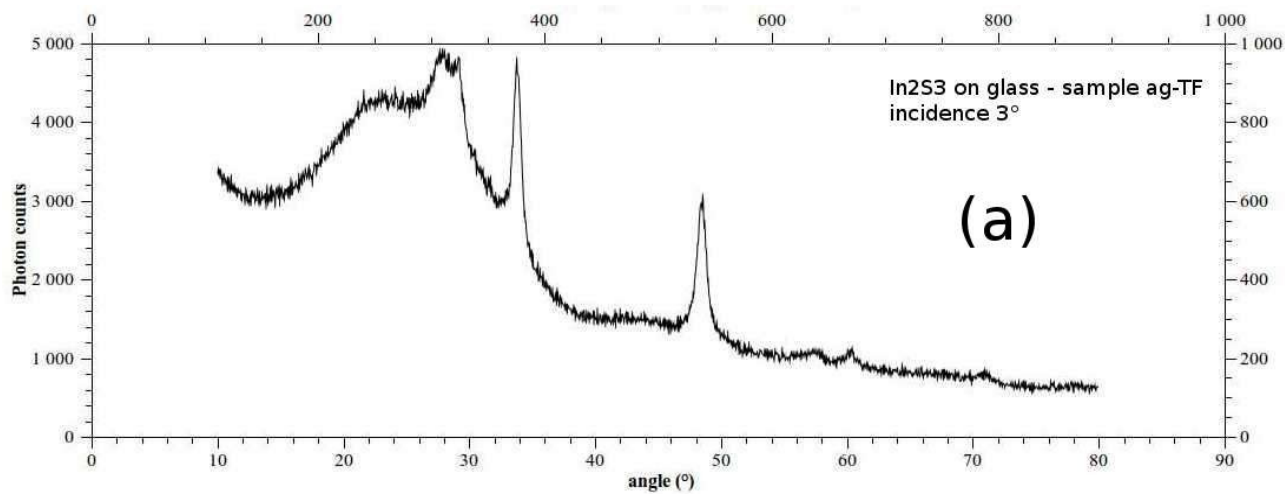

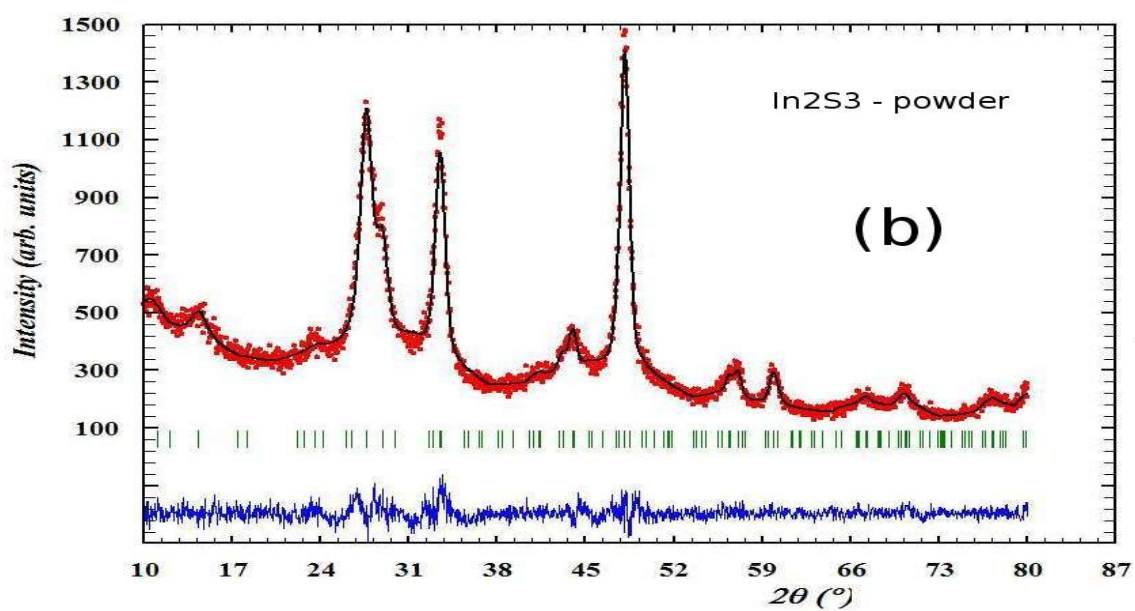

**Figure 1 : X ray diffraction pattern**

**(a) of an as-grown In$_2$S$_3$ film (« ag TF » sample)**

The background is mainly due to the glass substrate but could contain an amorphous component inherent to the layer. Comparing this pattern with JCPDS data No. 73-1366, all reflexions may be indexed, that allows to identify the compound as being indeed the β–In$_2$S$_3$.

**(b) of an as-grown In$_2$S$_3$ powder extracted from the film (« ag FP » sample)**

Dots represent experimental points, and the continuous line is the result of profile matching performed with Fullprof (see text). Vertical segments show the position of calculated diffraction peaks. The lower curve represents the difference between measured points and calculated curves.

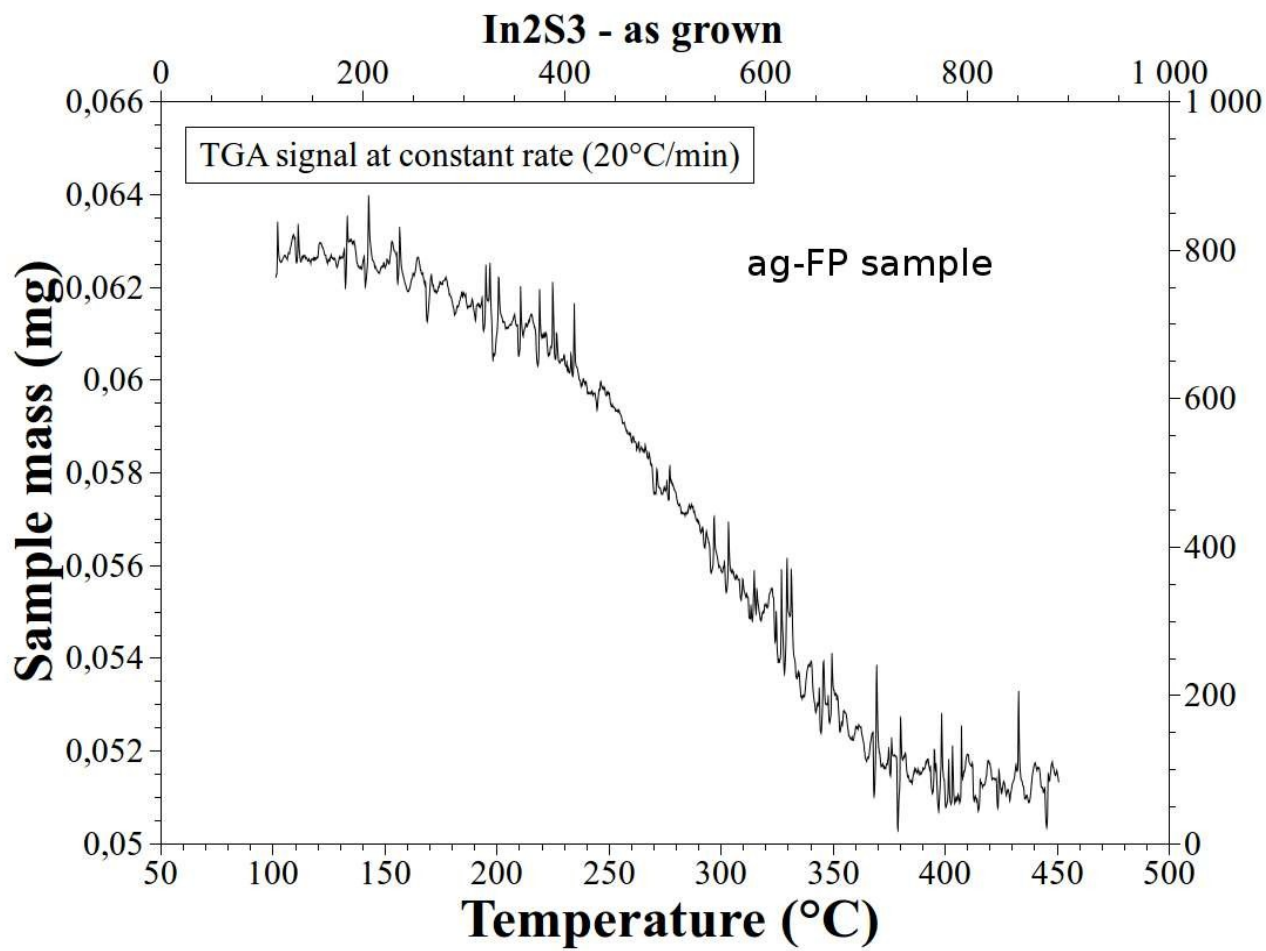

**Figure 2 : TGA result obtained with powder extracted from the as – grown film (« ag FP » sample) in air atmosphere**

The mass loss attributed to oxidation occurs above 150°C and extends in a rather wide temperature range till approximately 380°C, above which the mass remains constant.

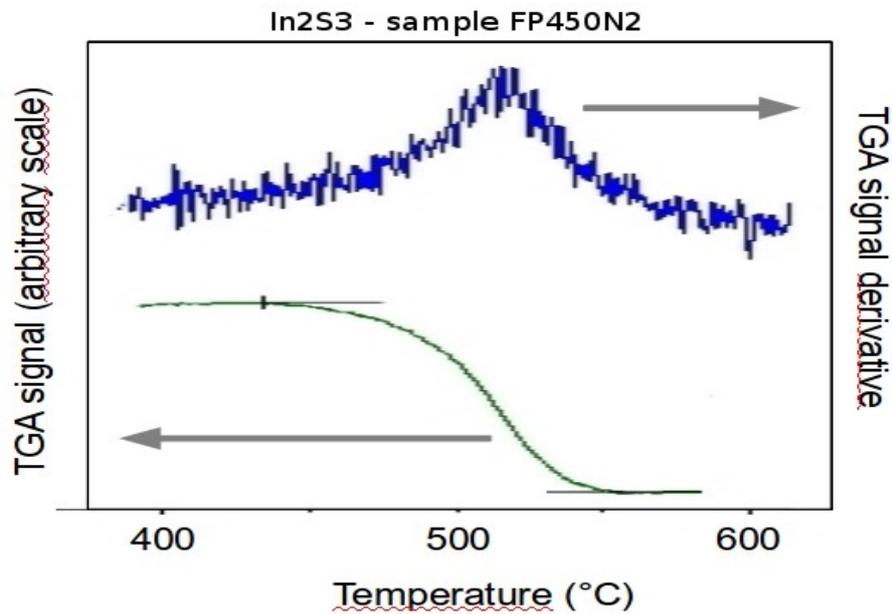

**Figure 3 : TGA result obtained in air atmosphere with the ag-film powder (see fig.2) after thermal annealing in an inert atmosphere ($N_2$) at 450°C**

In comparison with Fig.2, the inflection point is shifted towards high temperature (it is located around 520°C – see derivative maximum). The mass loss occurs in a smaller temperature interval (it begins in the neighbourhood of 440°C and vanishes around 550°C).

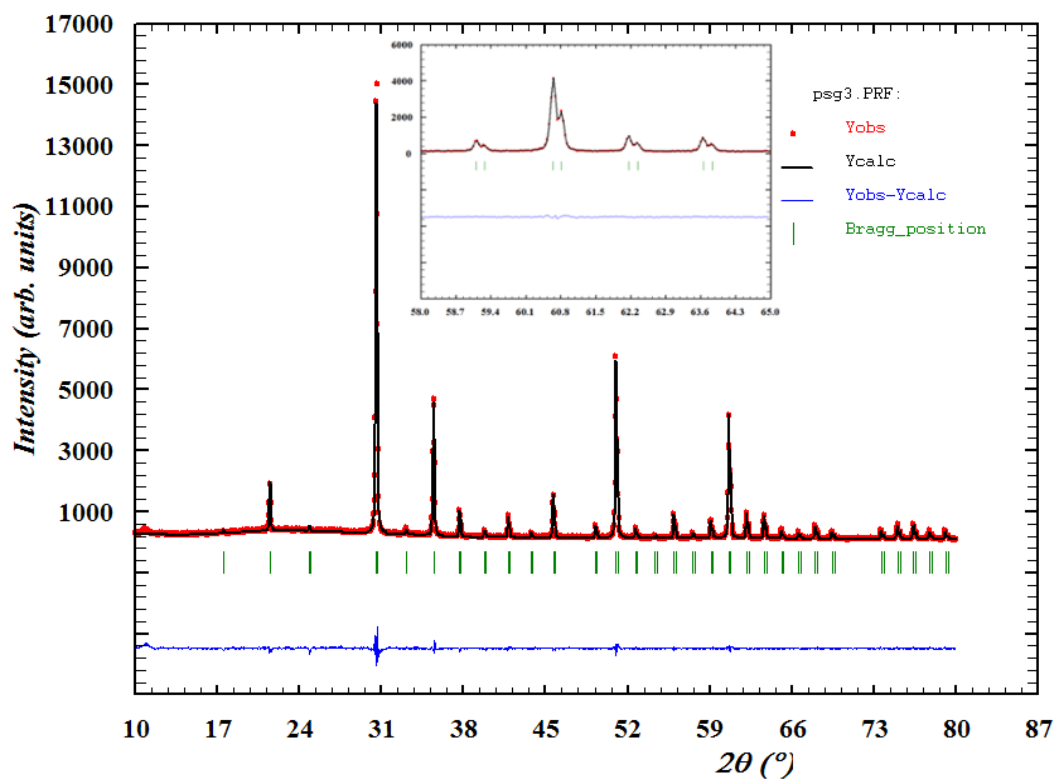

**Figure 4 : X ray diffraction pattern after annealing in air atmosphere at 450°C of ag-FP sample (see text)**

Comparing this pattern with JCPDS data confirms that the final product is $In_2O_3$ and that the mass loss observed in TGA measurements is associated with the substitution of S by O atoms. Dots correspond to experimental points, full line to the calculated profile resulting from profile matching, and vertical segments represent the position of calculated diffraction reflections. The lower curve is the difference between experimental points and calculated diffraction pattern. Inset shows a magnification of a particular region, that better allows to realize the high quality of the fit since all experimental points of diffraction pattern lie remarkably on the calculated curve.

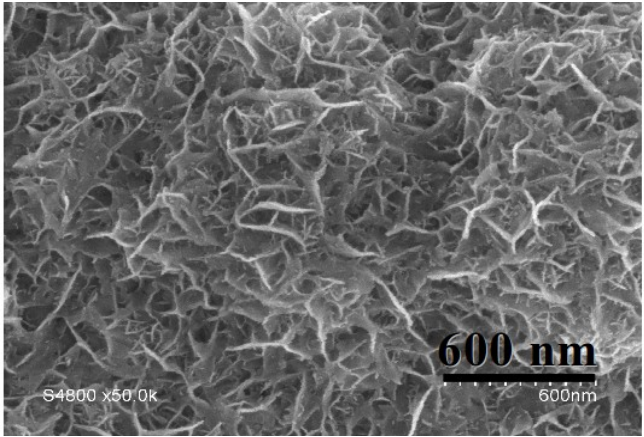 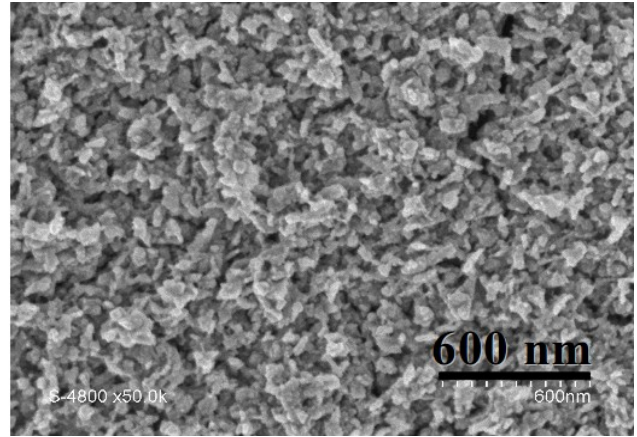

(a)          (b)

**Figure 5 : SEM pictures obtained on In$_2$S$_3$ films**

5a presents the texture of the as-grown film deposited on glass (sample ag-TF), and 5b of the same film after annealing in nitrogen atmosphere at 450°C (heating rate of 10°C/min and annealing at 450°C during 15 min- sample TF-450N2). The annealing leads to a coarsening of the texture.

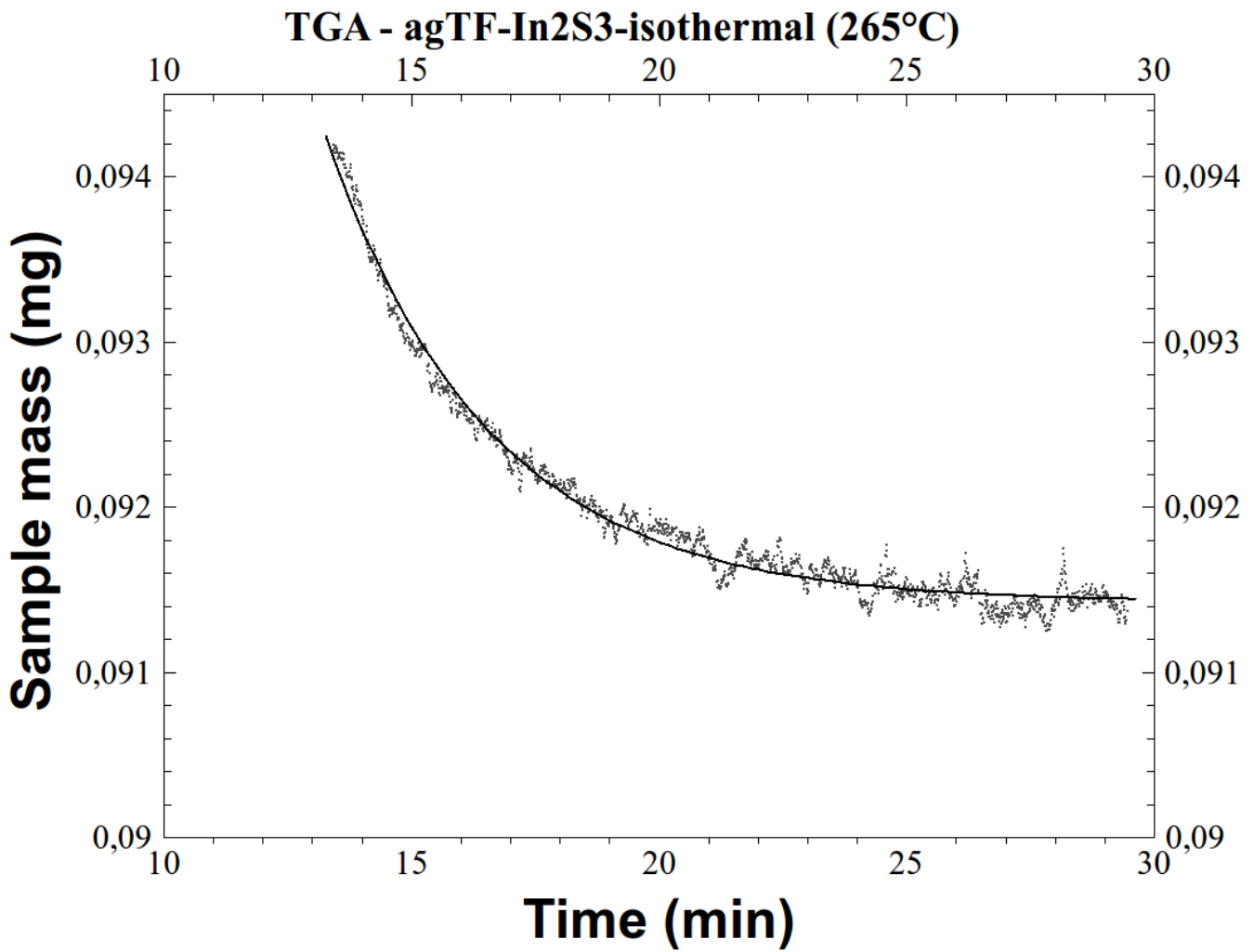

**Figure 6 : Isothermal TGA measurement on the same powder as in Fig.2 (« ag FP » sample) in air atmosphere**

After the heating run corresponding to Fig.2, temperature is stabilized in the region where the mass loss occurs – here at 265°C. We observe that the mass loss goes on as a function of time, and stabilizes at a value that corresponds to an uncomplete oxidation.

Dots represent experimental points and the full line the result of a fit according to

$m = m_o + A \exp[-(t-t_o)/\tau]$ that leads to $\tau = 3,30 \pm 2 \times 10^{-2}$ min

(Adjusted $R^2$ = 0,9856)